# Zoea – Composable Inductive Programming Without Limits

Edward McDaid FBCS
Chief Technology Officer
Zoea Ltd

Sarah McDaid PhD
Visiting Senior Research Fellow
London South Bank University

**Abstract**

Automatic generation of software from some form of specification has been a long standing goal of computer science research. To date successful results have been reported for the production of relatively small programs. This paper presents Zoea which is a simple programming language that allows software to be generated from a specification format that closely resembles a set of automated functional tests. Zoea incorporates a number of advances that enable it to generate software that is large enough to have commercial value. Zoea also allows programs to be composed to form still larger programs. As a result Zoea can be used to produce software of any size and complexity. An overview of the core Zoea language is provided together with a high level description of the symbolic AI based Zoea compiler.

## 1. Introduction

Automation has been a relentless force in software development throughout most of its history. This drive has accelerated in the last decade or so with approaches including test automation [1], continuous integration [2] and continuous deployment [3] becoming common. The benefits of automation are often expressed in terms of improved quality, greater development velocity and reduced time to market. From the business perspective many of these benefits translate into lower cost and greater efficiency.

Yet software development remains an expensive undertaking. For most organisations the greatest component in software development cost is direct labour. This is mainly due to the fact that coding is a labour intensive activity. It is also partly due to a significant and increasing global shortage of skilled developers. As a result any degree of automation in the coding activities of the software development lifecycle could have significant benefits.

At the same time only around 0.3% of the worlds population are software developers [4]. Given the significant and growing importance of software in everyday life and business this means that the vast majority of people are effectively disenfranchised, limited to being consumers of software provided by a small minority.

Many people think the world would be a better place if more people could create their own software. To this end there have been calls to teach programming more widely [5]. However, all mainstream programming languages are complex and require significant time and effort to learn. In addition, technology changes rapidly so ongoing effort would be required to maintain any skills so acquired. As a result it is difficult to imagine a significant percentage of people becoming software developers in the current sense of the term. An alternative strategy might be to enable people to some how generate the software they want.

The automatic generation of software from some kind of specification has been a long standing objective of computer science research. While significant and promising progress has been made in this field it has proved difficult to move beyond the generation of relatively small programs.

This paper describes a new inductive programming language called Zoea. Zoea is a very simple language that resembles set of automated test cases. Despite its simplicity Zoea is intended to be a general purpose language that can also be used to generate programs of any size.

## 2. Related work

Automatic Programming is a long standing if somewhat archaic umbrella term used to characterise a wide variety of code generation and transformation approaches [6]. In these approaches the input is usually either a program or a model of some form and the output is a modified or generated program. In terms of making software development easier approaches involving models as input are more attractive than those involving other programs.

Program synthesis is a sub-field of automatic programming in which the input is a model in the form of a mathematical specification of the required program [7]. In common with formal methods, programs generated through program synthesis may also have the benefit of being verified as correct with respect to the model. While program synthesis may be useful for the development of mission critical applications it is less obviously appropriate for casual end user programming in its current form.

Inductive programming is another sub-field of automatic programming in which the specification takes the form of a set of constraints, input-output examples and/or traces of program data during execution [8]. Interest in inductive programming has waxed and waned several times over the decades as it was applied in turn to different programming paradigms [9,10,11]. Inductive programming has an attractive operating model as it can require little or no end user programming knowledge. After nearly fifty years of research it is finally starting to make its way into mainstream IT products [12] although it is still capable of producing only relatively small programs.

In parallel with the above the domain of software development has been a natural subject for artificial intelligence research aimed at producing tools that codify software development knowledge [13]. More recently there has been significant interest in applying deep learning to enhance software development tools [14,15]. Keyword programming is an interesting approach that translates an unordered set of keyword symbols supplied by the user into one or more candidate program fragments [16]. This approach is based on prior extraction of code fragments and associated keywords from a corpus of existing software.

Visual programming is another broad area aimed at making software development more accessible through the use of diagrams [17]. The key challenge in this area remains that of formulating notations that are adequate for software specification without just using diagrams as a literal visual representation of an equivalent text based program [18].

Within the software industry there are many technologies that either automate the production of code or obviate the need for code that would otherwise be required. Such technologies usually operate within a single application tier (e.g. user interface or database) or across a specific interface between components. Generation of complete programs that are highly constrained (such as reports) has been possible for a long time. Attempts at producing more general software in this way tend to resemble selection from a set of pre-defined program configurations.

The widespread availability of spreadsheet software revolutionised the ability of many non-programmers to produce software albeit in a limited sense [19]. A notable characteristic of spreadsheets is the primary importance of data and its direct manipulation over processes, objects and other abstractions. The shallow learning curve has led to many millions of users with a basic level of expertise although advanced skills are much less common. While spreadsheets can meet some of the basic software needs of many businesses they are widely regarded as inappropriate for the development of sophisticated applications.

Test driven development is a common software development approach in which automated test cases are produced before the code to which they relate [20]. The test cases are intended to demonstrate the complete and correct operation of the software with respect to the requirements and as such they represent a sort of executable specification. The widespread use of test automation and the established role of test cases within the software development lifecycle makes them an appealing basis for an accessible specification language.

It is interesting that our understanding of what programming languages are and how they are used is still unfolding. Part of this insight comes from the availability of large corpora of existing software and the tools to analyse them. For example a recent study showed that the vast majority of control loops are small and have simple conditions while a few dozen standard patterns can account for most of the looping constructs used by developers [21]. Such findings appear to make the automatic generation of software a somewhat more tractable proposition.

## 3. Zoea

Zoea is a new programming language that produces software automatically from a specification that resembles a set of automated functional test cases. The overarching goal in producing Zoea was to make software development as simple as possible and this is reflected in the simplicity of the language. At the same time Zoea is intended to be a practical general purpose programming language.

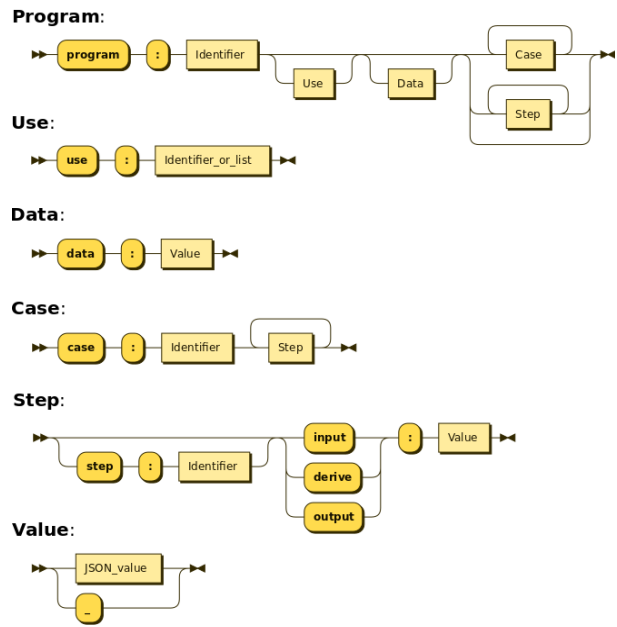

**Figure 1**

Zoea does not have many of the elements that are normally present in other programming languages such as variables, conditions, statements or control structures. Instead the user simply specifies the behaviour of the program they want by creating a set of input and output examples. Zoea is thus a declarative language.

The following sections provide a high level introduction to the core features of the Zoea language. This includes some simple examples of Zoea programs. These are not meant to demonstrate the limits of what is possible but rather they are intended to give a sense of how the language works and what it is like to use.

### 3.1 Terminology

In the context of Zoea an executable unit of software is called a **program**. Every program has a single **input** and a single **output**. Each program input and output is composed of a hierarchy of one or more scalar or composite data types. An **element** is any node in an input or output hierarchy. Each input and output has a single **root** element. Every element in an input or output is addressable via a **path** which is a possibly empty list of array indexes and/or keys. The path which is an empty list denotes the root element. Programs that require multiple inputs or multiple outputs can simulate these by convention using a list of elements. (This is similar to the way in which multiple values can exist within a single HTTP request or response.)

A **user** is a person that interacts with Zoea to create and execute one or more programs.

An **instruction** is a unit of computation from which programs are assembled A program is composed of zero or more instructions. Instruction inputs and outputs follow the same rules as program inputs and outputs Any program that successfully compiles and that is accessible to a user can be used as an instruction in another program.

A program is specified by a set of zero or more **test cases** (or simply **cases)**. Each case corresponds to a single linear **scenario** consisting of one or more **steps**. A step has one input, or one output, or one **derived** value. A derived value describes an internal data value that is intermediate between an input and an output. Test case input, output and derived values follow the same rules as program inputs and outputs.

### 3.2 Grammar

Listing 1 shows the grammar for the core elements of Zoea in EBNF notation. For simplicity Listing 1 omits definitions for JSON_Value [23], Identifier (number or string) and Identifier_or_list (Identifier or list thereof).

The same grammar is also shown in Figure 1 as syntax diagrams.

```
Program ::= "program" ":" Identifier
            Use ?
            Data ?
            ( Case Case * |
              Step Step * ) ?
Use   ::= "use" ":" Identifier_or_list
Data  ::= "data" ":" Value
Case  ::= "case" ":" Identifier
          Step Step *
Step  ::= ( "step" ":" Identifier ) ?
          ( "input" |
            "derive" |
            "output" ) ":" Value
Value ::= ( JSON_value | "_" )
```

**Listing 1**

### 3.3 Examples

Zoea code superficially resembles YAML [22] however Zoea does not rely on layout for structuring. Zoea programs consist of a set of tags and values that are separated by a colon character (':'). White space has no meaning and can occur in any quantity before and after tags and values. Tags and values do not need to be quoted unless they contain spaces.

Values in Zoea are basically JSON data [23] although as with YAML quotes are optional for strings that contain no spaces or escaped characters. A value can also be represented by a single unquoted underscore character. This means that the corresponding value is unspecified as it is not used in the current test case. This construct can be employed instead of an actual value in any input or output element. A quoted underscore or an underscore in any other context is a single character value.

One or more Zoea programs can be defined in a text file. The name of the text file is not significant and is only used to submit the programs it contains to the compiler.

All programs in Zoea must include an identifier which is the name of the program. A program name can be any non-empty string so long as it is unique across all programs for a given user.

The simplest program that is possible to create using Zoea is a null operation. Listing 2 shows a complete program that does nothing.

```
program: do_nothing
```

**Listing 2**

Inputs and outputs are specified using tags called input and output. Listing 3 is a program that outputs a greeting.

```
program: say_hello
  output: "hello world"
```

**Listing 3**

The user is free to lay out the program however they want. They could for example place everything on one line or have each tag and value on separate lines but neither of these approaches would be particularly usable.

In a program consisting of a single test case with an output and no input Zoea will assume that the output is a literal. This is one simple example of the many ways in which Zoea reasons about the test cases.

Input and output values can be numbers, strings or lists. Quotes can be omitted from strings if they are not required. When quotes are required then single or double quotes can be used.

Input and output tags always take a single value. If multiple inputs or outputs are required then a list is used as in Listing 4.

```
program: concatenate
  input: ["abc", "xyz"]
  output: 'abcxyz'
```

**Listing 4**

A Zoea program consists of any number of test cases and each case can have an identifier. The case identifier can be any value but each has to be unique within a particular program. Cases are introduced by the case tag. If a program has only a single case then the case tag can be omitted as in Listing 4. Listing 5 includes three test cases.

```
program: median
  case: 1 input: [3,5,6] output: 5
  case: 2 input: [1,2,4,6,9] output: 4
  case: 3 input: [2,4,5,8] output: 4.5
```

**Listing 5**

The median is the middle value of a sorted list of numbers or the average of the middle two numbers if the length of the list is even. When a program is defined using multiple test cases the order in which the test cases are defined is not important.

It should be clear that some tags (program, case and step) remain in scope until another tag of the same name is encountered. For example everything after a program tag relates to the same program until another program tag occurs. Similarly the same case remains in scope until the next case or program tag.

Inputs and outputs with multiple elements can be accomplished using lists. Listing 6 calculates the minimum and maximum values in a list of numbers.

In this example two cases are required in order to avoid any ambiguity. Using just case 1 might be interpreted as 'select the second and fourth elements' or even 'select the elements with an odd index'. While using case 2 on its own could have meant 'sort' or 'reverse list'. This shows that it is still important to test the programs that Zoea generates and it may be necessary to modify the cases or add more of them to achieve the desired result.

```
program: min_and_max
  case: 1
    input: [7,3,11,15,6]
    output: [3,15]
  case: 2
    input: [2,1]
    output: [1,2]
```

**Listing 6**

Values can be embedded in input or output strings and these will be extracted or interpolated as appropriate. Listing 7 counts the words and characters in a string. Zoea will determine that the numeric values in the output are calculated while the rest of the output is composed of literal string values.

```
program: count_words_and_chars
  input: 'how now brown cow'
  output: '4 words and 17 characters'
```

**Listing 7**

The next example is a little more complex. Listing 8 extracts fields from a delimited record to produce a JSON object and in the process it carries out some simple data transformations.

```
program: parse_record
  input: "001 SMITH JOHN 07/24/79 UK"
  output: {
            "name": "John Smith",
            "date": "24-07-1979"
          }
```

**Listing 8**

The output name field is assembled from two input fields and the text case is changed. The format of the date field is also changed.

Conditionals are an important element of any programming language. Listing 9 selects only the females from a list of people and their corresponding genders.

```
program: select_females
  input: [
          [fred,    male],
          [wilma,   female],
          [barney,  male],
          [betty,   female]
         ]
  output: [wilma, betty]
```

**Listing 9**

Zoea includes knowledge about a range of software development techniques and one of these is regular expressions. Listing 10 shows a program that extracts a number of different value types from an input string.

```
program: extract_data
  input: 'xyz21-07-1969abc123pqr22.7'
  output: ['21-07-1969', 123, 22.7]
```

**Listing 10**

What is interesting about this example is that Zoea is able to determine that a regular expression is an appropriate strategy for solving this problem and then goes on to produce the correct regular expression required.

At the same time Zoea would also produce other candidate solutions using other instructions such as string operations. In this case the regular expression version is selected as the best solution by virtue of being the simplest.

Derived values are one of the mechanisms that allow Zoea to produce more complex programs. This can be used to create a linear sequence or directed graph of intermediate values between input and output steps in a test case. Listing 11 shows a very simple example of how this mechanism could be used to capitalise the first letter of the words in a sentence.

```
program: ini_caps
  input: 'how now brown cow'
  derive: [how, now, brown, cow]
  derive: ['How','Now','Brown','Cow']
  output: 'How Now Brown Cow'
```

**Listing 11**

As it happens Zoea is easily capable of producing this program directly without the need for derive steps. However they can be useful for more complex programs and they also to allow solutions to be found more quickly. Derive can be essential when the output depends on a condition that is expressed in terms of values that do not themselves appear in the output.

Up to this point only test cases with a single input and/or a single output have been considered. Listing 12 shows a case that consists solely of a number of outputs.

```
program: factorial
  output: [0,1]
  output: [1,1]
  output: [2,2]
  output: [3,6]
```

**Listing 12**

Zoea recognises this as a sequence and the generated program will continue to run beyond the initial values in the test case.

```
program: sales_tax
  case: 1
    input: 1000
    output: 175
  case: 2
    input: 2000
    output: 350

program: price_including_tax
  use: sales_tax
  input: 1000
  output: 1175
```

**Listing 13**

An important mechanism in Zoea is the ability to form complex programs by combining a number of simpler programs. This is called composition. In practise existing Zoea programs are used as additional instructions in the new program. Composition supports a style of bottom up incremental development. In order to be able to use an existing program in a new program the existing program must have compiled successfully. The developer indicates that an existing program should be used in a new program by including a 'use' tag in the new program. The value associated with this tag is either the name of a single program or a list of program names. Listing 13 shows a simple example of composition in Zoea.

Programs often require reference data that is relatively static and which can be large. It would be tedious to have to provide information like this as additional inputs for every test case. Instead Zoea provides a facility to specify such values once at the start of a program. Listing 14 shows an example of this feature in use.

```
program: is_week_day
  data: [monday,tuesday,wednesday,
         thursday,friday,
         saturday,sunday]
  case:1 input: thursday output: true
  case:2 input: 'MONDAY' output: true
  case:3 input: banana output: false
  case:4 input: '' output: false
```

**Listing 14**

This paper presents a range of simple examples of Zoea programs. Most of these require very little explanation due to the simplicity of the language. The examples also cover virtually all of the syntax of the language as it currently exists.

### 3.4 Language enhancements

The language is currently being extended to allow the production of programs that are sensitive to date and time or that use random values. Similar facilities will be used to enable programs to query or manipulate the environment. An ability to organise programs into packages is also planned.

Currently Zoea programs are written in a text file and submitted to the compiler using command line tools. In the future a simple graphical user interface will be developed that will serve as an interactive development environment. It is expected that this will increase the usability considerably.

The provision of user feedback is recognised as an important aspect of the development process although very little has been done in this regard to date. At present the user submits a Zoea program to the compiler and after some time is notified whether compilation was successful or not. In formulating a solution Zoea can make a number of assumptions regarding the test case data and its transformation. It would be useful to make these assumptions visible to the user in a convenient format and also provide a mechanism that would allow the user to accept or reject them. Such interaction would make the development process more engaging and probably also more effective.

Another planned enhancement is the ability to describe the side effects of programs. For example it should be possible to specify how a particular test case might change the contents of one or more database tables or files. While such facilities will be supported in the source language it will be more convenient for users to specify such behaviour through a user interface. This will also enable inputs and outputs to be expressed in terms of user interaction allowing Zoea to be used for the development of GUI applications.

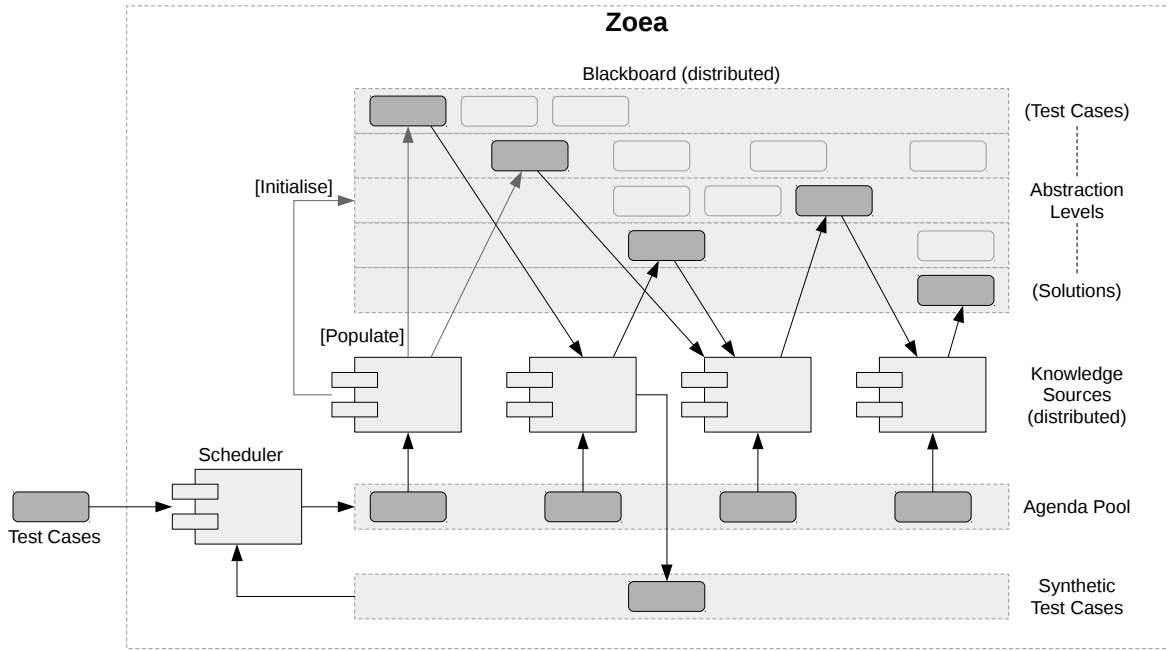

**Figure 2**

Zoea currently targets its own virtual machine. It would also be possible to target other virtual machines or to generate code in other programming languages. This feature is not a high priority relative to those above.

## 4. Zoea compiler

Zoea consists of the programming language together with an associated compiler and virtual machine. The Zoea compiler uses symbolic artificial intelligence to transform test cases into software. It does this by applying a number of reasoning strategies in conjunction with a wide range of software development knowledge.

It is quite common for systems analysts to communicate some or even all the requirements for a program as a set of test cases. As a result turning a set of test cases into a program is a daily activity for many software developers. Nevertheless as with many of the things that people accomplish habitually it is easier to do than it is to describe or explain.

### 4.1 Problem space

In the most general terms the Zoea compiler needs to generate a program that will transform a specific input into the correct output for each of a number of test cases. In this sense an acceptable solution needs to work with all of the test cases.

At the same time a partial solution may work with a subset of the test cases, or with a subset of the output elements, or with both of these limitations. Partial solutions can be combined in various ways to form complete solutions. For example the identification of more than one partial solution across different test cases can be interpreted as the existence of top level conditional logic within the solution.

For any given set of test cases there are an infinite number of possible solution candidates that can be identified. One or more of these solution candidates may correspond to the behaviour that the user is trying to describe. These actual solutions will vary in terms of their size and complexity. As a general rule Zoea should find the simplest solution that works with all of the test cases.

Another subset of solution candidates have the characteristic that for a given test case input they always produce the literal corresponding test case output. Indeed there is a degenerate subset of candidate solutions that always produce the literal

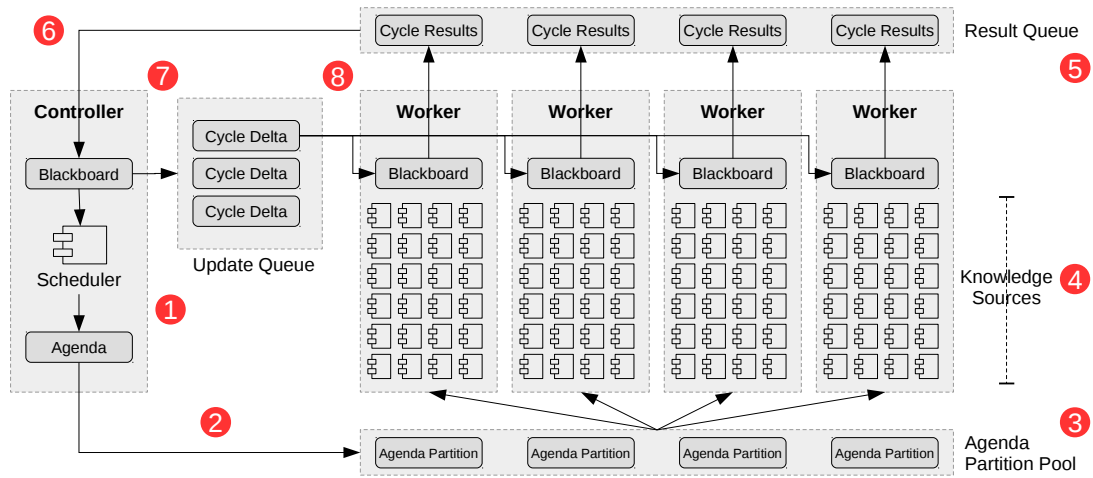

**Figure 3**

corresponding output for every specified test case. This may or may not be the required behaviour.

On the other hand the required program may be expected to work with a range of inputs of which the test cases represent only a sample. Again there is no way of knowing whether this is the case or not. As a result solution candidates must be evaluated and ranked so that the most appropriate member can be identified. The following dimensions are available:

1. Simple – complex;
2. Complete – composite;
3. General – specific.

Other factors being equal a simple, complete or general solution will be preferred over the alternatives. However, things become more complex when solution candidates vary across multiple dimensions. For example a more complex solution may be preferred if it is also complete.

The length of time required to find a solution is also important but not in the sense of a comparative metric. In the interests of usability Zoea must minimise or at least limit the amount of time spent in trying to find one or more candidate solutions. If an acceptable solution has not been found in the available time then the user should be able to choose whether to invest more time to find better candidates or alternatively provide additional or better test cases.

### 4.2 Architecture

The software architecture of the Zoea compiler is a variant of the blackboard model [24]. This comprises a standard blackboard data structure organised as an abstraction hierarchy, together with a number of knowledge sources and a relatively simple scheduler. The Zoea blackboard is also recursive and distributed.

Figure 2 provides a simplified view of the Zoea architecture. This shows the single scheduler, some knowledge sources and the blackboard data structure. In practise there are many more knowledge sources than are shown. Test cases are submitted to the compiler and the data they contain is used to initialise the corresponding abstraction levels in the blackboard. The scheduler identifies and selects agenda items that are distributed to the knowledge sources through a shared agenda pool. The knowledge sources update various abstraction levels in the blackboard to build up different fragments of the solution until a complete solution is obtained. Some knowledge sources also create new synthetic test cases that correspond to sub-problems that need to be solved or hypotheses that need to be tested. These synthetic test cases are handled in the same way as user originated test cases effectively making the blackboard recursive. All of the knowledge sources are themselves embarrassingly parallelisable.

The technical architecture consists of a cluster of homogeneous nodes with a single threaded process per physical core. There is a single controller node which is home to the scheduler and the master copy of the blackboard, and many worker nodes (currently 32).

Each worker node has an eventually complete copy of the blackboard and all knowledge sources. Cluster nodes communicate using persistent message queues. Figure 3 shows a more detailed

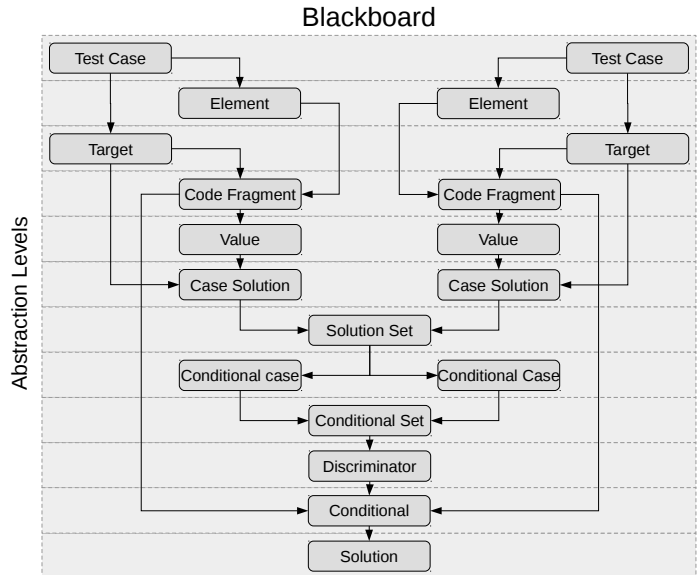

**Figure 4**

view of the deployment architecture and the blackboard lifecycle.

The blackboard operates in cycles. Within each cycle a single knowledge source is selected by the scheduler and all activity on all workers is undertaken with respect to that knowledge source.

The scheduler maintains a single agenda which it derives from the initial or current state of the blackboard. Agenda entries relating to the selected knowledge source are partitioned into work packages (tasks) and placed in the agenda pool. Each worker claims a task and invokes the relevant knowledge source with the subset of the agenda it contains.

On completion of each task the worker collects the set of changes to the blackboard and posts these back to the controller via the results queue. The worker then continues by claiming another task if more are available.

As the cycle progresses the controller asynchronously collects results from workers and updates its copy of the blackboard. At the end of the cycle the controller creates a consolidated delta that uniquely captures all blackboard changes in the current cycle. This is sent to all workers so they can update their local copies of the blackboard. At this point all of the blackboards are again identical.

The coordination activities including handling of results and updates accounts for less than 2% of available computational resources.

### 4.3 Abstraction hierarchy

The abstraction hierarchy models and integrates the key concepts used by the various knowledge sources in formulating code solutions that correspond to test cases. The abstraction levels represent a general progression from the test cases through available and derived values to partial and complete solutions. The abstraction levels include:

- test cases;
- input and output elements;
- derived values (symbolic and numeric);
- code fragments;
- target values;
- case solutions;
- case set solutions;
- program solutions;
- solution code.

The data on the blackboard represents a set of more or less promising solution fragments at different stages of identification, characterisation and elaboration. It is worth noting that progression from test cases to solution code is not a strictly linear process. Instead knowledge sources respond to changes at one or more specific abstraction levels to produce, enhance or remove elements on different levels. The blackboard model allows this to happen in more or less any order. This characteristic of blackboard systems is often referred to as the opportunistic application of knowledge.

The knowledge sources effectively act as a set of experts that embody software development knowledge and/or apply various reasoning strategies. All interaction between knowledge sources takes place via the blackboard and in this sense they are said to be loosely coupled.

Most of the knowledge sources represent different aspects of software development knowledge such as string manipulation, conditionals, collections and so on. Other knowledge sources are more abstract such as those that reason about combinations of sets of solutions.

Each knowledge source identifies opportunities where its knowledge can be applied. This often corresponds to the creation of assumptions which ultimately may turn out to be valid or invalid. In any given scenario preference may be given to the veracity or otherwise of a given assumption but Zoea ensures that both alternatives are explored if necessary. Policy regarding specific assumption types are one example of the many kinds of heuristics employed in the compiler.

Figure 4 shows a simplified view of the blackboard abstraction levels corresponding to a simple conditional statement. At the top are entries relating to the test cases. The arrows represent associations between entries in different abstraction levels. The test cases are connected to (input) elements and (output) targets. Code fragments combine elements to form new intermediate values of which only a couple are shown - although in reality there are likely to be many. Case solutions represent candidate solutions to individual test cases and these are combined to form sets that represent possible candidate solutions. In this case there is no single solution that produces all of the test cases so a conditional is required.

The conditional logic is determined through a similar process that ultimately produces a composite conditional referencing the previously determined code fragments. Any alternative solutions would coexist on the relevant abstraction levels. Ultimately any judgement regarding the relative fitness of solutions will be conducted with respect to the ensemble.

At any point in time the solution islands that exist across different blackboard abstraction levels together with the hierarchy of recursive blackboard instances corresponding to synthetic test cases form the solution space. There are many options with respect to how this solution space can be explored. As a hedge Zoea uses multiple strategies concurrently with the winner being either:

- the first consistent solution to all test cases or
- the optimum solution available after a given period of time.

The blackboard architecture makes it easy to integrate a variety of problem solving strategies without having to worry about the order in which they are invoked or even how they interface with one another.

Pattern recognition applied to the input and output can yield important clues about how the intended program needs to work. A simple example is where an output is composed of different elements some of which are directly available in the input. In this case the real problem now becomes how to assemble the output elements that are not directly available from the input. In this scenario the initial problem is reformulated as a synthetic test case that tries to determine how the unknown values are assembled from the input.

Another area where direct inspection can make a significant difference is in formulation of the instruction set. If it was known what specific instructions are required in order to generate the solution then the task would become significantly easier. Zoea uses pattern recognition to produce a candidate solution instruction set based on the input and output type signatures and other characteristics. This approach is often but not always successful. Where it does work the solution is generally obtained much more quickly. In the small proportion of cases where it is less successful the alternative strategy of using the complete instruction set is also pursued so the net result is a relatively small amount of wasted effort.

Part of the sophistication of the Zoea compiler comes from the blackboard being used recursively. Normally the compiler operates on test cases provided by a user to produce a complete program. This process often involves the creation of many hypotheses in various knowledge sources. A convenient way to minimise the need for state management is to translate such hypotheses into sets of synthetic test cases. These synthetic test cases which are generally simpler are then processed by new instances of the blackboard.

A single set of test cases can result in a hierarchy of synthetic test cases that represent different assumptions about the solution. All of the information relating to each instance of the blackboard is completely partitioned from all other instances.

Although Zoea is characterised as 'inductive programming' this really reflects the operating model more than the technological approach. Induction is not that important in Zoea and more use is made of pattern recognition and abductive reasoning. Similarly state space search is avoided whenever possible but it is often necessary so considerable effort has been put into finding good heuristics.

### 4.4 Enabling technologies

Zoea incorporates a number of technologies that enable it to produce larger programs without the need for composition or explicit statement of derived values. A few of these are discussed here.

One of the biggest problems in generating programs automatically is the size of the search space. A naive approach might simply take the input values and combine them with the available instructions to produce intermediate values, and repeat this process until the required output value is obtained. However, the search space rapidly becomes enormous if the program is longer than three or four instructions.

Zoea avoids the use of state space search whenever possible. This is mainly accomplished by applying pattern recognition to the test case data to identify mapping and transformation relationships between input and output elements. Pattern recognition is also used to customise the instruction set. Nevertheless there are situations where search is required and Zoea uses a number of techniques to make this more efficient.

For any given search space, the space of distinct intermediate values is significantly smaller than the space of all possible intermediate value derivations. This is true because there are many different ways in which the same intermediate value can be derived. As intermediate values are produced from other intermediate values the number of derivations grows rapidly with increasing depth. Zoea reduces the size of the search space significantly by decoupling the tracking of intermediate values from how they are are derived. When an interesting value is found a reverse search is performed to find its derivations. On its own this is a powerful technique but it can also be extended considerably.

The state space that is composed of test case and intermediate values, and instructions can be thought of as a single (enormous) data flow graph. This graph is the output of the distinct value approach described above. Within this graph each candidate solution exists as a data flow subgraph. When searching a graph it is much easier to find a linear data flow sequence between two given values than it is to find a complete subgraph. Such data flow sequences often contain unrecognised intermediate values. In a data flow subgraph for a candidate solution some of these unrecognised values correspond to (symbolic or numeric) deltas with respect to known values. This correspondence can be used in turn to 'grow' the branches of the subgraph from the nodes of the linear data flow sequence.

When programs are generated according to a language grammar there are many programs that are syntactically correct but semantically nonsensical for a variety of reasons. For example a program can be written that first reverses a list and then sorts it. In this example the reverse operation is redundant as its side effects are completely obliterated by the subsequent sort operation. Zoea has functionality that enables it to generate only candidate programs that are sensible as well as syntactically correct.

For each program there exists an infinite number of functionally equivalent programs of which one or more will be the smallest in terms of the number of instructions they contain. (This is intuitively obvious as redundant instructions can be added to any program to produce another functionally equivalent program.) Zoea avoids considering duplicate programs by generating only programs and fragments of programs that are functionally distinct.

Programming languages with large instruction sets tend to produce smaller programs and consequently have shallower search spaces than languages with a small number of instructions. This is a frequent observation concerning concatenative languages such as Forth. The Zoea virtual machine has a large and open ended instruction set which makes compiled Zoea programs much smaller than they otherwise would be.

Zoea was developed from first principles using a clean room process and without reference to any existing or reported system, technology or approach. It contains no third party components and not a single line of third party code.

### 5. Discussion

Inductive programming is a technology enabled process that involves one or more people articulating software requirements which are transformed into code. For any non-trivial application it is natural that this process should be incremental and iterative. The design of Zoea recognises the importance of these characteristics by allowing more complex programs to be developed using stepwise-refinement and assembled through composition. This is similar to how many professional software developers prefer to work but it does not require any greater level of technical expertise.

There was a lot of interest in the use of input-output examples in the early days of inductive programming research. However a criticism that became widespread was that this approach would not scale as more complex programs would require very large numbers of examples to adequately describe complex conditional logic. It is worth remembering that this view dated from an era when people still wrote monolithic programs consisting of tens of thousands of lines of code in a single file. Even so this viewpoint turns out to have been an incorrect assessment. This is because the same problem had already been solved in the context of decision tables through hierarchical decomposition. These days it is unremarkable for large code bases consisting of millions of lines of code to have an average method length in single digits. It is also now commonly accepted that complex behaviour and large applications are amenable to test automation.

Zoea is not intended to replace existing programming languages and there are probably some areas of software development where this approach may not be particularly appropriate. Nevertheless there are many areas where this approach could be safely used and also justified. Much of the software that is currently developed is not complex but rather repetitive and formulaic. Zoea is well suited for this kind of development - potentially freeing up professional software developers for more interesting work. Zoea could also allow many more people who aren't professional programmers to create software to meet their own requirements.

Zoea is a very different kind of programming language and as such it is difficult to predict how languages of this kind will be used in practise and how they might evolve. For this reason Zoea as it currently exists should be viewed as a first iteration rather than the finished product.

## 6. Conclusions

Zoea is a very simple programming language that bares a strong resemblance to a set of test cases. It can be learned quickly and can be used to produce a wide range of software. The ability to form larger programs through composition means that it could enable people with no previous software development expertise to produce moderately complex software with little training. Zoea is made possible by the symbolic AI and blackboard architecture in its compiler.

## Acknowledgements

This work was supported entirely by Zoea Ltd (https://www.zoea.co.uk). All trademarks mentioned in this paper are the property of their respective owners.

Syntax diagrams were produced using Railroad Diagram Generator (https://www.bottlecaps.de/rr/ui/).

## References

[1] Dustin, E., Rashka, J., Paul J. (1999) Automated Software Testing. New York, Addison Wesley. ISBN 978-0-201-43287-9.

[2] Beck K. (1999) Embracing Change with Extreme Programming. IEEE Computer 32 (10), 70-77.

[3] Holmstrom Olsson, H., Alahyari, H., Bosch, J. (2012) Climbing the "Stairway to Heaven" -- A Multiple-Case Study Exploring Barriers in the Transition from Agile Development towards Continuous Deployment of Software. Proceedings of the 2012 38th Euromicro Conference on Software Engineering and Advanced Applications. IEEE Computer Society, 392–399.

[4] Evans Data Corporation (2019) Global Developer Population and Demographic Study 2019 Volume 1. Evans Data Corporation. Available from: https://evansdata.com/reports/viewRelease.php?reportID=9 (Retrieved: 1 Nov 2019).

[5] Andreessen M. (2011) Why Software Is Eating The World. Wall Street Journal August 20, 2011. Available from: https://www.wsj.com/articles/SB10001424053111903480904576512250915629460 (Retrieved: 1 Nov 2019).

[6] Rich, C., Waters, R.C. (1993) Approaches to automatic programming. Advances in Computers 37, 1-57. Boston, Academic Press.

[7] Manna, Z., Waldinger, R. (1980) A deductive approach to program synthesis. ACM Transactions on Programming Languages and Systems. 2 (1), 90–121.

[8] Kitzelmann, E. (2010) Inductive programming: A survey of program synthesis techniques. Approaches and Applications of Inductive

Programming. Lecture Notes in Computer Science 5812, 50–73. Berlin, Springer-Verlag.

[9] Summers, P.D. (1977) A methodology for LISP program construction from examples. Journal of the ACM. 24 (1), 161–175.

[10] Shapiro, E.Y. (1981) The model inference system. Proceedings of the 7th international joint conference on Artificial intelligence 2, 1064-1064. San Francisco, Morgan Kaufmann.

[11] Katayama, S. (2008) Efficient exhaustive generation of functional programs using Monte-Carlo search with iterative deepening. PRICAI 2008: Trends in Artificial Intelligence, 199-210. Berlin, Springer-Verlag.

[12] Galwani, S., Hernandez-Orallo, J., Kitzelmann, E., Muggleton, S.H., Schmid, U., Zorn, B. (2015). Inductive Programming Meets the Real World. Communications of the ACM 58 (11), 90–99.

[13] Rich, C., Waters, R.C. (1988) The Programmer's Apprentice: A Research Overview. IEEE Computer 21(11), 10-25.

[14] Vincent, J. (2019) This ai-powered autocompletion software is gmail's smart compose for coders – upgrading codingautocompleter tools with deep learning, Available from: https://www.theverge.com/2019/7/24/20708542/coding-autocompleter-deep-tabnine-ai-deep-learning-smart-compose (Retrieved 1 Nov 2019).

[15] Kite (2019) Kite announces intelligent snippets for python. Available from: https://dev.to/kite/kite-announces-intelligent-snippets-for-python-13i0 (Retrieved: 1 Nov 2019).

[16] Little, G., Miller R. (2009) Keyword programming in Java. Automated Software Engineering 16 (1). 37-71.

[17] Myers, B.A. (1990) Taxonomies of visual programming and program visualization. Journal of Visual Languages & Computing 1 (1), 97-123.

[18] Whitley, K.N. (1997) Visual Programming Languages and the Empirical Evidence For and Against. Journal of Visual Languages & Computing 8(1), 9-142.

[19] Gislason, H. (2018) Excel vs. Google Sheets usage — nature and numbers. Available from: https://medium.com/grid-spreadsheets-run-the-world/excel-vs-google-sheets-usage-nature-and-numbers-9dfa5d1cadbd (Retrieved: 1 Nov 2019).

[20] Beck, K. (2002). Test-Driven Development by Example. Vaseem: Addison Wesley. ISBN 978-0-321-14653-3.

[21] Allamanis, M., Barr, E.T., Bird, C,, Devanbu, P., Marron, M., Sutton, C. (2018) Mining semantic loop idioms. IEEE Trans Software Engineering 44 (7), 651-668.

[22] Ben-Kiki, O., Evans, C., Döt Net, I. (2009) YAML Ain't Markup Language (YAML™) Version 1.2. Available from: https://yaml.org/spec/1.2/spec.html (Retrieved: 1 Nov 2019).

[23] ECMA International (2017) The JSON Data Interchange Syntax. ECMA-404. 2nd edition. Available from: http://www.ecma-international.org/publications/files/ECMA-ST/ECMA-404.pdf (Retrieved: 1 Nov 2019).

[24] Nii H.P. (1986) The Blackboard Model of Problem Solving and the Evolution of Blackboard Architectures. AI Magazine 7 (2), 38-53.